\documentclass[lettersize,journal]{IEEEtran}
\usepackage{amsmath,amsfonts,amssymb}
\usepackage{array,pifont}
\usepackage[caption=false,font=normalsize,labelfont=sf,textfont=sf]{subfig}
\usepackage{textcomp}
\usepackage{stfloats}
\usepackage{url}
\usepackage{verbatim}
\usepackage{graphicx}
\usepackage{multirow}
\usepackage{cite}
\usepackage{threeparttable}
\usepackage{tikz}
\usetikzlibrary{fit, calc}
\usetikzlibrary{decorations}
\usepackage{verbatim}
\usepackage{pgfplotstable}
\usepackage[noend]{algpseudocode}
\usepackage{listings}
\usepackage{xcolor}

\begin{document}

\title{Securing Operating Systems Through Fine-grained Kernel Access Limitation for IoT Systems}

\author{Dongyang Zhan,~\IEEEmembership{Member,~IEEE, }
        Zhaofeng Yu,
        Xiangzhan Yu,
        Hongli Zhang,
        Lin Ye,
        and~Likun Liu*
\IEEEcompsocitemizethanks{\IEEEcompsocthanksitem D. Zhan, Z. Yu, X. Yu, H. Zhang, L. Ye, L. Liu are with the School of Cyberspace Science, Harbin Institute of Technology, Harbin,
Heilongjiang, 150001.\protect\\
E-mail: \{zhandy, yuxiangzhan, zhanghongli, hityelin, liulikun\}@hit.edu.cn
\IEEEcompsocthanksitem * Corresponding Author: liulikun@hit.edu.cn}
}

\maketitle

\begin{abstract}
With the development of Internet of Things (IoT), it is gaining a lot of attention. It is important to secure the embedded systems with low overhead. The Linux Seccomp is widely used by developers to secure the kernels by blocking the access of unused syscalls, which introduces less overhead. However, there are no systematic Seccomp configuration approaches for IoT applications without the help of developers. In addition, the existing Seccomp configuration approaches are coarse-grained, which cannot analyze and limit the syscall arguments. In this paper, a novel static dependent syscall analysis approach for embedded applications is proposed, which can obtain all of the possible dependent syscalls and the corresponding arguments of the target applications. So, a fine-grained kernel access limitation can be performed for the IoT applications. To this end, the mappings between dynamic library APIs and syscalls according with their arguments are built, by analyzing the control flow graphs and the data dependency relationships of the dynamic libraries. To the best of our knowledge, this is the first work to generate the fine-grained Seccomp profile for embedded applications. 
\end{abstract}


\section{Introduction}\label{sec:introduction}
\IEEEPARstart{I}{nternet} of Things (IoT) is developing rapidly and has become one of the most important computing platforms. IoT devices have been used in a lot of environments, such as industrial control/sensing systems, home automation, etc. However, the security risks are raised with the IoT development. The IoT/embedded systems have strict requirements for resource occupation and energy consumption, so the security problem is more serious\cite{liao2020security,li2018towards}. And, the performance and resource requirements of security tools running inside the embedded devices are also very strict\cite{yaqoob2019internet}. 

Securing embedded operating systems is essential for IoT security\cite{zikria2019internet}. There are many vulnerabilities \cite{CVE-2017-7308,CVE-2017-5123,CVE-2016-8655} in OS kernels that can be  exploited to perform the privilege escalation attacks from applications. After the privilege escalation, attackers can control the device and access all of the data. Therefore, securing the embedded operating system and isolating the IoT applications are important. 

Fortunately, the Linux Seccomp can be used to secure the operating systems. According to the Eclipse IoT Developer Survey report, Linux is the most popular OS for IoT devices\cite{iot-survey}. Seccomp is a security module of the Linux kernels, which can block the unused syscalls from user-space applications. Since the vulnerabilities in the blocked syscalls are isolated, the attack surface is reduced. Seccomp is an embedding module in the Linux kernel, which usually introduces low overhead to the whole system, so it is meaningful to secure the embedded systems based on Seccomp\cite{xu2016toward}. 

By using Seccomp, the kernel attack surface can be reduced by preventing the unused kernel syscalls that may contain exploitable vulnerabilities from being exploited by user-space applications. However, it is challenging to generate Seccomp configuration systematically. To analyze the dependent syscalls of a binary, the dynamic tracking and static analysis approaches are proposed for x86 binaries. The dynamic tracking approaches collects the invoked syscalls of the target binaries dynamically. But it cannot obtain the complete dependent syscall list, which is not acceptable in practical scenarios. The static analysis approaches generate the mapping between the dependent library APIs and syscalls, so that the syscall list can be obtained. But, these systems are not designed for embedded systems. Furthermore, the existing static and dynamic analysis approaches are coarse-grained, which cannot analyze the mapping between library API (e.g., fopen, flose) and syscall arguments. Limiting syscall arguments is essential for system security, because some arguments affect the control flows and the functions of the syscalls. 

In this paper, a systematic Seccomp configuration analysis approach is proposed to reduce the attack surface for embedded devices (e.g., routers), which can not only block the unused syscalls but also limit the arguments of the allowed syscalls. The dependent syscalls and the corresponding arguments are obtained by analyzing the target binaries and the dependent libraries. Firstly, the target binaries are analyzed to obtain the invoked library APIs and the corresponding arguments. Then, the mapping between library APIs and syscalls is built by analyzing the dependent libraries. To further limit the syscall arguments, the relationship between API arguments and syscall arguments is analyzed. Finally, the fine-grained Seccomp configuration can be generated.

To generate the mapping between library API arguments and syscall arguments, a systematic data dependency analysis approach is proposed, which can construct a data dependency graph for every library API argument. The graph is constructed by control flow graph analysis, static taint analysis and symbolic execution. Firstly, the control flow graph of every library API is constructed, so that the mapping between library APIs and syscalls can be obtained. Then, the static taint analysis is employed to track the propagation of every argument. Static taint analysis and dynamic data tracking are two kinds of data flow analysis methods. Since we aim to find out the comprehensive data dependency relationship, we use the static approach in this paper. During the taint analysis and symbolic execution, the variables that influence the data propagation are analyzed to find out if the syscall arguments are determined based on the library API arguments. If so, the syscall arguments can be determined.

To the best of our knowledge, this is the first work to perform the fine-grained syscall access control for the embedded systems. By analyzing the target applications and the dependent libraries, the types and arguments of the system calls can be limited at best effort. Therefore, the embedded system can be secured with very low overhead by leveraging Seccomp. 

In summary, the contributions of our paper are as follows.
\begin{itemize}
  \item A systematic attack surface reduction approach based on Seccomp is proposed to secure the embedded systems by limiting the accessible syscalls and the corresponding arguments of the embedded applications.
  \item The applications and the dependent libraries are analyzed statically to find out the comprehensive dependent syscalls, so that the corresponding Seccomp configuration can be generated.
  \item To further reduce the attack surface, a data dependency analysis approach is proposed, which constructs the data dependency graph of every syscall argument to generate the mapping between API arguments and syscall arguments.
\end{itemize}

The rest of this paper is organized as follows. The background and key insight are described in Section \ref{background}. Section \ref{s:design} presents the system design. Section \ref{s:implementation} gives some implementation details of our system. Section \ref{s:evaluation} evaluates our system. The related work is summarized in Section \ref{s:related-work}. Finally, Section \ref{s:conclusion} concludes this paper.  

\section{Background \& Key Insight} \label{background}
\subsection{Linux Seccomp}
There is a computer security facility in the Linux kernel called Seccomp (secure computing mode)~\cite{Seccomp}, which allows processes to use Berkeley Packet Filter (BPF) to filter syscalls through configurable policies. The Linux kernel exposes many syscalls to user space. In recent years, many discovered vulnerabilities are related to Linux syscalls, and some of these vulnerabilities (e.g., CVE-2017-7308, CVE-2017-5123) may be used to damage the operating system kernel. It can be seen that the syscalls exposed to user space introduce a large attack surface. Considering that the user space process only uses a subset of the syscall set provided by the Linux, we can narrow the attack surface by reducing the set of syscalls accessible by the user-space processes. And, Seccomp provides a method for the processes to specify the set of syscalls they can call. Seccomp is typically configured in two ways. First, an application can configure the Seccomp policy by using several syscalls. Second, an application can first configure the Seccomp and then execute another application that should be protected, so the executed application will also be limited with the same Seccomp policy. When Seccomp is configured, if the process tries to access a syscall that is not allowed, the operating system will reject the call. The Linux Seccomp is widely used by lots of well-known programs, such as OpenSSH, etc. However, many developers do not specifically provide Seccomp configuration files, so this may expose the operating system to certain risks. If an unrestricted program that provides Internet services is controlled by an attacker in some way, the vulnerabilities in all accessible syscalls may be exploited. Therefore, strengthening the security of the operating system through Seccomp is very important for the maintainer of the system. This paper aims to narrow the attack surface by precisely restricting the syscalls that the program can access without requiring additional work from the developers.

\subsection{Shrinking Kernel Attack Surface}
Shrinking the kernel attack surface is one of the most important approaches to secure the systems\cite{ghavamnia2020confine}, based on the observation that the fewer system calls an application can access, the less chance it can exploit the kernel vulnerabilities\cite{nabla}. There are some approaches to reduce the attack surface in different scenarios. 

Recompiling the programs with the library OS\cite{tsai2014cooperation} to reduce the dependent syscalls is a possible way. The Nabla Container \cite{nabla} integrates most of the syscalls into the container images by redesigning the applications and recompiling them with the library OS. After that, only 7 syscalls can be accessed by the containers. The experimental results show that the accessed code of the tested Nabla Container is less than that of Docker containers. But, not all applications can be recompiled, so the idea of Nabla Container is not widely applied by industry.

Seccomp is widely used to reduce the available syscalls for applications. By using Seccomp flexibly, SPEAKER\cite{lei2017speaker} can apply different control policies to an application according to its execution status (e.g., initialization and servicing). SPEAKER collects the dependent syscall sets by tracking the target applications dynamically. The dynamic syscall collection approaches are also employed by \cite{wan2017mining,barlev2016secure,niu2020sasak}, which can generate Seccomp policies for applications based on the invoked syscalls. However, the dynamic syscall collection cannot obtain the comprehensive dependent syscalls because it is difficult for dynamic execution to cover all of the execution paths. 

To overcome the problems of dynamic syscall collection, some static dependent syscall analysis approaches are proposed. Confine\cite{ghavamnia2020confine} is designed for containers, which builds the mapping between library APIs and syscalls based on the static analysis of the library source code. After obtaining the dependent library APIs, Confine can generate Seccomp configuration for containers. Unlike Confine, Sysfilter\cite{demarinis2020sysfilter} and \cite{zeng2014tailored} builds the mapping between library APIs and syscalls based on the static analysis of library binaries. Chestnut\cite{canella2021automating} analyzes the dependent syscalls through a static compiler-based source code analyzer and a binary analyzer, and it can restrict the set of allowed syscalls dynamically. However, these approaches cannot analyze the arguments of the dependent syscalls, making the access control coarse-grained. In addition, these approaches are designed for x86 programs, and no system is designed for embedded applications. There are several challenges to apply the approaches for x86 programs to ARM programs, due to some special designs for the ARM platform. First, the syscall mechanism of the ARM platform is different from that of x86. Second, the inline assembly code used to invoke syscalls in glibc under the ARM platform is different from that under the x86 platform, so the binary analysis should be redesigned for the ARM platform. Furthermore, the instruction analysis and the data propagation analysis of ARM applications are different from those of x86 applications. In this paper, we aim to propose a fine-grained syscall limitation approach to limit both syscalls and their parameters precisely for embedded IoT applications.

\subsection{Threat Model}

Our approach aims to reduce the kernel attack surface by preventing the unused kernel syscalls that may contain exploitable vulnerabilities from being exploited by user-space applications. We assume that the attacker can hijack the control flow of user-space applications (or call the target applications) to invoke some syscalls that contain vulnerabilities. Before the attacks, the kernel is secure and not controlled by attackers.

We assume the source code of popular dependent libraries is available, especially for glibc, which is open-source and usually used by programs or other dynamic libraries to invoke syscalls on different platforms (e.g., x86 and ARM). For the self-written libraries and the programs that use inline assembly code to invoke syscalls, we analyze the binaries of them. In order to increase the generality of our system, we analyze the binaries of the target programs.

\subsection{Observation}

For Linux-based systems, the dependent syscalls of a program can be obtained by analyzing the dependent libraries. The user-space programs usually invoke syscalls through dynamically-linked libraries. A program can rely on many libraries (e.g., libssl, glibc, etc.). Among them, the glibc is one of most important one, since it is usually used by programs or other libraries to invoke syscalls. In addition, most self-written libraries also leverage the glibc to invoke syscalls. Since glibc is open-source, we can obtain the source code of it. Based on the analysis of the code, we can build the mapping between the library APIs and syscalls. Based on the dependent API list of a target program, the dependent syscalls can be obtained.

To perform the dependent syscall analysis for IoT applications, the target binary should be firstly analyzed to obtain the dependent libraries and the corresponding APIs. For Linux, most of dependent libraries (e.g., glibc and other basic libraries) are open-source, so it is possible to analyze the mapping based on the open-source libraries. In some systems, some dependent libraries can even be replaced. There are two scenarios of applying the Seccomp-based security tools in embedded systems. When the application can be modified by the security administrators, the generated Seccomp configuration can be embedded in the application through the configuration syscalls. If the application cannot be modified, the application can be executed by another programmable application and Seccomp is configured in the programmable application. So that, the application is also limited.

Seccomp is able to filter the syscall arguments (i.e., flags), which can further secure the system. Syscall arguments include many types, such as flags and pointers, and most of them are passed from the library API parameters. But, the syscall arguments that Seccomp can filter are flags. Seccomp cannot filter points. But, currently there is no automatic argument analysis tool to enrich Seccomp configuration with argument policies. Filtering syscall arguments is important for system security, because many attacks\cite{mulliner2015breaking,url-file} rely on some critical syscalls with special arguments. For instance, some backdoor attacks\cite{url-file} tamper with the original login file of the target server. If the application cannot invoke the open syscall with the write flag, these attacks can be blocked. 

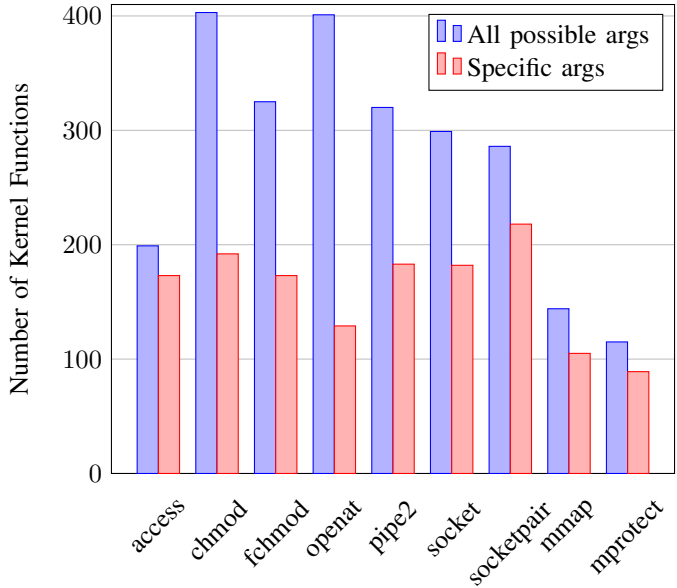
\begin{figure}
    \centering
    \begin{tikzpicture}
    \begin{axis}[
    width  = 0.5\textwidth,
    major x tick style = transparent,
    ybar=0.05pt,
    bar width=8pt,
    ymajorgrids = true,
    ylabel={Number of Kernel Functions},
    symbolic x coords={access, chmod, fchmod, openat, pipe2, socket, socketpair, mmap, mprotect},
    xtick = data,
    xticklabel style={rotate=45},
    scaled y ticks = false,
    ymin=0,ymax=410,
    ytick style={draw=none},
    legend cell align=left,
    legend style={
    },
    ]
    \addplot table[x=cert,y=union] {\dispdata};
    \addplot table[x=cert,y=intersection] {\dispdata};
    \legend{All possible args, Specific args}
    \end{axis}
    \end{tikzpicture}
    \caption{Comparison of invoked kernel functions of syscalls with different arguments.
    }
    \label{fig:comperison-func}
\end{figure}

Besides, limiting syscall arguments can reduce the number of accessible kernel functions of a syscall, which can reduce the possibility of triggering kernel vulnerabilities. We select several syscalls (e.g., socket, access, fchmod, etc.) to do the experiment by tracking the invoked kernel functions with different syscall arguments (or flags). In the experiment, we use trace-cmd to track and record the kernel functions accessed by each syscall given different arguments, which results are shown in Figure \ref{fig:comperison-func}. The blue bar shows the number of executed kernel functions of a syscall with all possible arguments, and the red bar shows the minimal function number with a specific argument. From the figure of 9 examples, we can find that limiting the syscall arguments can significantly reduce the possible accessed kernel functions. Taking the socket syscall as an example, we take AF\_INET, AF\_INET6, and AF\_UNIX for the first argument, and SOCK\_STREAM, SOCK\_DGRAM, and SOCK\_RAW for the second argument. In this way, a total of nine argument usage situations are formed. As shown in Table \ref{table:kernel_function_number}, when the value of the argument is not restricted, we tracked 299 executed kernel functions. When argument 1 is restricted to AF\_INET, AF\_INET6, and AF\_UNIX, the number of executed kernel functions is 267, 252, and 214, respectively. Limiting only one argument can greatly reduce the number of kernel functions accessed by a syscall. When the remaining arguments are further restricted, we can further reduce the number of accessed kernel functions. From Table \ref{table:kernel_function_number}, we can see that the results of mmap are similar with those of socket. Based on the assumption that the more accessible code, the greater the probability of exploiting the vulnerabilities\cite{nabla}, further limiting syscall arguments is essential for system security.

\begin{table*}[!t]
  \caption{Numbers of executed kernel functions with different limited syscall arguments.}
  \label{table:kernel_function_number}
  \centering
  \begin{tabular}{cccccc}
    \hline
    \textbf{syscall}&\textbf{Total Num}&\textbf{Arg1}&\textbf{Num2}&\textbf{Arg2}&\textbf{Num3} \\
    \hline
    \multirow{9}{*}{socket}&\multirow{9}{*}{299}
    &\multirow{3}{*}{AF\_INET}&\multirow{3}{*}{267}&SOCK\_STREAM&211\\
    &&&&SOCK\_DGRAM&234\\
    &&&&SOCK\_RAW&252\\
    \cline{3-6}
    &&\multirow{3}{*}{AF\_INET6}&\multirow{3}{*}{252}&SOCK\_STREAM&221\\
    &&&&SOCK\_DGRAM&196\\
    &&&&SOCK\_RAW&227\\
    \cline{3-6}
    &&\multirow{3}{*}{AF\_UNIX}&\multirow{3}{*}{214}&SOCK\_STREAM&211\\
    &&&&SOCK\_DGRAM&207\\
    &&&&SOCK\_RAW&206\\
    \hline
    \multirow{9}{*}{mmap}&\multirow{9}{*}{144}
    &\multirow{3}{*}{PROT\_EXEC}&\multirow{3}{*}{120}&MAP\_SHARED&118\\
    &&&&MAP\_PRIVATE&120\\
    \cline{3-6}
    &&\multirow{3}{*}{PROT\_READ}&\multirow{3}{*}{117}&MAP\_SHARED&114\\
    &&&&MAP\_PRIVATE&117\\
    \cline{3-6}
    &&\multirow{3}{*}{PROT\_WRITE}&\multirow{3}{*}{138}&MAP\_SHARED&134\\
    &&&&MAP\_PRIVATE&118\\
    \cline{3-6}
    &&\multirow{3}{*}{PROT\_NONE}&\multirow{3}{*}{116}&MAP\_SHARED&107\\
    &&&&MAP\_PRIVATE&116\\
    \hline
  \end{tabular}
  \footnotesize
 \\
 {Total Num: the number of invoked kernel functions with all possible arguments; Arg1: the first argument of the syscall; Num2: the number of invoked kernel functions with the first argument fixed; Arg2: the second argument of the syscall; Num3: the number of invoked kernel functions with the first and second arguments fixed.}
\end{table*}

\pgfplotstableread[row sep=\\,col sep=&]{
    cert    &   intersection   &   union    \\
    access	&	173 &   199 \\
    chmod	&	192	&	403 \\
    fchmod	&	173	&	325	\\
    openat	&	129	&	401	\\
    pipe2	&	183	&	320 \\
    socket	&	182	&	299	\\
    socketpair	&	218	&	286	\\
    mmap	&	105	&	144	\\
    mprotect	&	89	&	115	\\
}\dispdata

Fortunately, analyzing the possible dependent API arguments of programs can be performed by \cite{mishra2018shredder}. Based on the idea of it, it is possible to analyze the dependent API arguments of IoT applications. So, the focus of this paper is to further construct the mapping between API arguments and syscall arguments, so that Seccomp can be used to secure the embedded IoT system with low overhead.

\section{System Design}\label{s:design}
\subsection{System Overview}
\begin{figure*}
\centering
\includegraphics[width=0.85\textwidth]{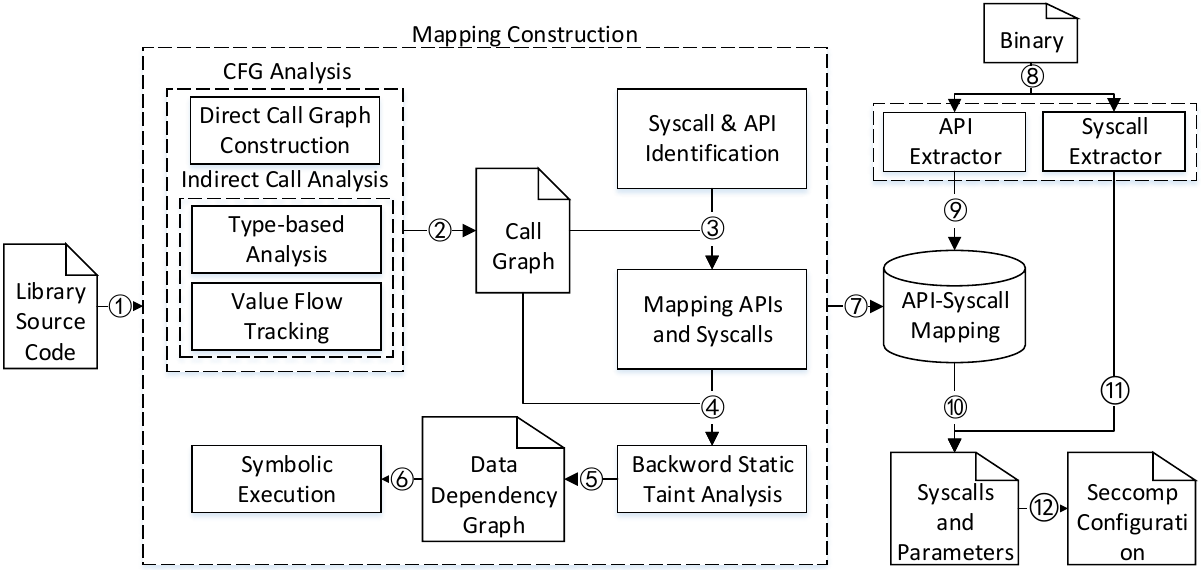}
\caption{The workflow of our system. \textcircled{1} The library source code is fed into the mapping construction module. \textcircled{2} The CFG Analysis module outputs the function call graph of the library. \textcircled{3} By identifying the syscall and API points in the call graph, the mapping between APIs and syscalls can be constructed. \textcircled{4} The backward taint analysis is used to find out the data sources of each syscall argument. \textcircled{5} The data dependency graph is constructed during the taint analysis. \textcircled{6} The symbolic execution is employed to analyze how the conditional judgment statements can affect the arguments in the data dependency graph. \textcircled{7} Based on the results of taint analysis and symbolic execution, the mapping between API arguments and syscall arguments is constructed. \textcircled{8} The target binary is analyzed to extract the dependent API and arguments. \textcircled{9} The extracted API and arguments are used to search the mapping \textcircled{10} for possible dependent syscalls and arguments. \textcircled{11} If the binary does not use a dependent library to invoke syscalls, the syscalls and arguments are extracted. \textcircled{12} Seccomp configuration with allowed syscalls and arguments can be generated.} \label{fig:arch} 
\end{figure*}

Our system does not block a list of vulnerable syscalls/arguments, but only allows necessary syscalls/arguments for programs to execute. So that, the vulnerabilities caused by blocked syscalls/arguments can be mitigated. The reason we do not block only vulnerable syscalls/arguments is that the vulnerable syscalls we get may be incomplete. Our system identifies the list of necessary syscalls/arguments for a program to execute, and blocks other syscalls/arguments. So, the vulnerabilities caused by all of the unnecessary syscalls/arguments will be mitigated. As a result, our system minimizes the OS attack surface.

There are two main stages in our system: the mapping construction stage and the binary analysis stage, which workflow is shown in Figure \ref{fig:arch}. The mapping construction analyzes the dependent libraries to construct the mapping between library APIs and syscalls, which also includes the mapping between API arguments and syscall arguments. The binary analysis extracts the dependent library APIs and the directly invoked syscalls according with the arguments of APIs and syscalls. Finally, the fine-grained Seccomp configuration can be generated based on the mapping and the dependent APIs.

The mapping construction stage aims to build the mapping between library APIs and syscalls and the arguments of them. To that end, the control flow graph of the dependent library is first analyzed. By searching the control flow graph, we can find the related syscalls of every library API. The control flow graphs of dependent libraries are analyzed by two steps, including the direct call graph analysis and the indirect call graph analysis, which are inspired by Confine\cite{ghavamnia2020confine} and \cite{lu2019detecting}. We design a compiler-based function call graph construction approach to construct the direct call graph. As we know, the glibc cannot be compiled by the LLVM/Clang compiler. So, we leverage the outputs of the GCC compiler, which can be generated during the compilation. A two-level indirect call analysis is employed to find out the possible targets of the indirect calls. It first leverages the type-based analysis to find out the possible address-taken callees. After that, a precise type-based callee analysis based on value flow tracking is employed to reduce the possible targets of indirect calls.

After constructing the control flow graph, the mapping between API arguments and syscall arguments is analyzed. Firstly, the data dependency graph is constructed by leveraging the backward taint analysis to find out the data sources of each syscall argument. If all of the data sources are determined (e.g., constants or API arguments), the possible value set of the argument can be analyzed. Our taint analysis starts from the syscall arguments and finds all of the data sources to construct the data dependency graph. During the analysis, all of the conditional judgment statements are extracted and added in the graph. So that the variables that affect the value of the syscall argument can be obtained. To find out the final argument value set, the symbolic execution is employed to analyze how the conditional judgment statements can affect the arguments in the corresponding code blocks or functions. Finally, the mapping between API arguments and syscall arguments is constructed.

Next, the binary analysis is employed to determine the dependent library APIs and arguments of a target binary. The invocations of library APIs can be obtained by disassembling the target binary. We adopt the API argument analysis approach of \cite{mishra2018shredder} to obtain the arguments of the dependent APIs at best effort. Based on the mapping and dependent APIs, the dependent syscalls and arguments can be obtained. Finally, Seccomp configuration with allowed syscalls and arguments can be generated.

In Linux environments, most binaries invoke syscalls through dependent libraries, but some binaries invoke syscalls through the syscall() API, or embedded assembly code directly. When a binary invokes syscalls through the assembly  instructions (i.e., the syscall instructions), the analysis approach of \cite{mishra2018shredder} can be employed to get the syscall number and the corresponding arguments, which analyzes the syscall arguments through binary analysis. For the cases of invoking syscalls through the syscall() API, the proposed method of \cite{mishra2018shredder} can be used to find out the syscall name and arguments. If the API invocation is performed through an indirect call, our system analyzes it in a best effort way, since it is an opening problem to analyze indirect calls in binaries. The target API destination is searched backward from the API callsite using the method of \cite{mishra2018shredder}. If the API target cannot be obtained, the binary cannot be protected, and we leave this problem in the future work. Fortunately, this case is not common in our target binaries, and we did not encounter this case in the evaluation.

\subsection{Mapping APIs with Syscalls}\label{S:static-analysis}
This stage aims to construct the mapping between library APIs and syscalls by constructing the control flow graphs of the dependent libraries. To that end, the direct calls are analyzed with the assistance of the compiler and the indirect calls are analyzed by a two-level analysis approach. Based on the full control flow graph, the mapping can be constructed.

\subsubsection{Direct Call Graph Construction}
The LLVM compiler is widely used to generate the function call graphs of open-source software at compiling time. However, LLVM does not support the glibc compilation, so we need to use some other methods. Intuitively, we can construct the function call relationships from the C source files. But, C source code can be very complicated and it is difficult to parse them. On one hand, macros are often used in C source files. When parsing, it is inevitable to replace macros. There are many kinds of macros and their structures are complex. Furthermore, they are often accompanied by deep nesting. On the other hand, there are many functions with alias names in the C source code, which can cause many difficulties for our analysis. Therefore, how to analyze the source code automatically is a challenge. 

Another problem of the source code analysis is that the analysis of the syscall invocation is library-specific. Some syscalls are invoked through assembly code in the source file, so only analyzing the C code is not enough. As described in the glibc wiki\cite{glibc-wiki}, some syscalls are invoked through an assembly wrapper, and the compiler will directly convert a function name into the assembly code that invokes the syscall. So, the analyzer cannot find the syscall instructions of these syscalls in the source code.

To address these challenges, we build a direct call graph based on the disassembly of the glibc library. In the disassembly file, we can obtain the instruction information contained in each function. When the callsite is executed, its operand may be an immediate value or a register. If the operand is an immediate value, the immediate value is the offset of the function in the binary file, and the function name of the called function is enclosed by ``$<>$" at the end of the line. According to this rule, we can easily know which functions are directly called by a function. However, when the operand contains registers (register+immediate or only registers), it is an indirect call, which is analyzed later.

\subsubsection{Indirect Call Analysis}
The indirect call analysis is an opening problem in static analysis. In this paper, we adopt the two-layer indirect call analysis of \cite{lu2019does,lu2019detecting} to find the over-approximated callees of the indirect calls. The two-layer analysis first finds all of the address-taken functions. The address-taken functions are the functions which addresses are stored in global variables or objects. These addresses are usually used as the targets of indirect calls. Then, the type-based alias analysis\cite{niu2014modular,tice2014enforcing,farkhani2018effectiveness} can be used to identify the possible callees, by comparing the return types and arguments types between the callsites and callees. This approach can only find possible callees in a coarse-grained manner. We leverage the fine-grained type-based analysis approach\cite{lu2019does} to further reduce the possible callees. When an indirect callsite fetches a function address from an object, only the functions which addresses are stored in the same object type can be the possible callee. Based on this observation, we collect the detailed information of the objects that store the address-taken functions. The possible callees can be obtained by matching the corresponding object type. 

\subsubsection{Generating the Mapping}
The comprehensive control flow graph can be constructed by combing the direct call graphs and the indirect call graphs.

After constructing the control flow graph, the functions that invoke syscalls should be identified. There are three ways for the glibc to invoke syscalls\cite{glibc-wiki}. The first method is to wrap the API invocation to the syscall invocation, and the wrapper is in the syscall-template.S. After the compilation, the corresponding APIs will be implemented using the assembly code with different syscall numbers. The list of such APIs is kept in the source code of the glibc. The second way is to invoke syscalls through macro functions, such as ``SYSCALL\_CANCEL", which are also defined in the source code. The third method is to leverage the assembly code. Some functions in the glibc are programmed with the assembly code instead of the C language. The assembly code usually invokes syscalls. To determine if the functions is syscall-related, the assembly analysis is employed.

Next, the API function are identified in the control flow graph, which names are collected from the glibc document\cite{glibc-wiki}. By searching the control flow graph by starting with the API functions, the reachable syscalls of the API are collected. So that the API-Syscall mapping is constructed. However, this mapping is coarse-grained, which does not contain the argument relationship between APIs and syscalls. 

For the binaries that invoke syscalls through the swi instructions, a binary analysis approach is proposed. There are two ways to pass the syscall number. The first way is to pass the syscall number by the swi instruction itself, so the number can be obtained through analyzing the instruction. The other way is to pass the syscall number based on the r7 register, which can be analyzed by using the static data flow analysis. The collected syscalls will be added to Seccomp configuration.

\subsection{Mapping Arguments Between APIs and Syscalls}\label{s:dynamic}
To further limit the arguments of syscalls, the dependent API arguments are firstly identified based on the static analysis. Then, the mapping between API and syscall arguments is generated. The analysis includes two steps: the backward static taint analysis and the symbolic execution. The fine-grained Seccomp configuration with syscalls and arguments can be generated based on the mapping and the dependent API arguments of the target program.

\subsubsection{Extracting API arguments}
\begin{figure*}
\centering
\includegraphics[width=0.9\textwidth]{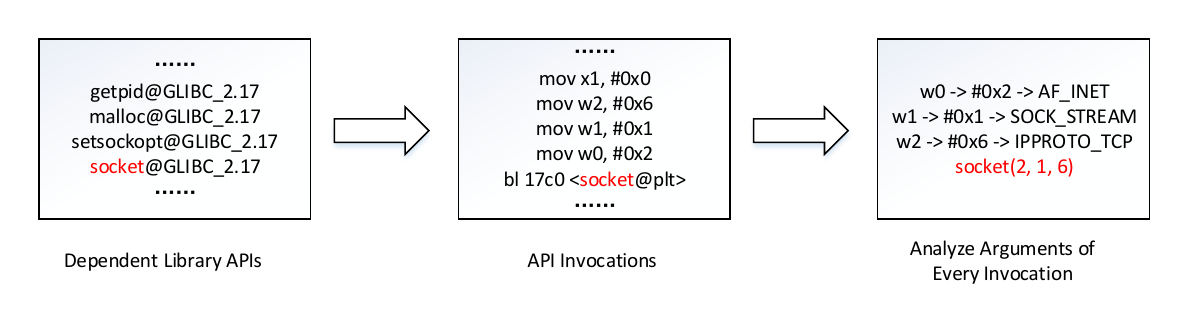}
\caption{An example to show the workflow of API argument extraction. In the example, the socket API was found to be used for syscall invocation. By analyzing the API invocation callsite, the w0-w2 were used for storing the API arguments. After analyzing the values of w0-w2, the API arguments (2,1,6) were determined.}\label{fig:api-para}
\end{figure*}

We adopt the backward data flow analysis of \cite{mishra2018shredder} to obtain the API arguments of applications statically, which workflow is shown in Figure \ref{fig:api-para}. It first pinpoints all of the API callsites in the target binary and then leverages the backward data flow analysis to extract the corresponding arguments at best effort. 

Our system is designed for the ARM Linux platform, so there are some parts specially designed for the ARM platform. First, the syscall mechanism of the ARM platform is different from that of x86\_64. In the x86\_64 environment, the relationship between the syscall number and the syscall name can be found in ``syscall\_64.tbl". However, there is no such file for the ARM64 environment, so "unistd.h" is used to establish a corresponding relationship. Taking the ``read" syscall as an example, the syscall number is 0 in the x86\_64 environment, while the syscall number is 63 in the ARM64 environment. Second, the inline assembly code used to invoke syscalls in glibc under the ARM platform is different from that under the x86\_64 platform, so the binary analysis should be designed for the ARM platform. For example, the inline assembly code to invoke syscalls in x86\_64 is "syscall", while the assembly code is ``svc 0" in ARM64. In addition, some functions in glibc are implemented in assembly code, which causes the analysis to be different under different architectures. Furthermore, the instruction analysis and the propagation analysis of ARM applications are different from those of x86\_64 applications.

As illustrated in Figure \ref{fig:api-para}, the binary analysis first extracts all of the callsites to the dependent library APIs. Data flow analysis is an opening problem in static binary analysis \cite{meng2016binary}, so the system works in a best-effort way.

The analysis targets at three types of variables, which are: 1) constant vales, 2) stack values and 3) register values. The backward analysis starts from the API arguments and tracks the value propagation through registers and memory. When the analysis reaches the beginning of a function, the parents of it in the control flow graph are continued to be analyzed. After the analysis, the result of an argument is either known and unknown. The unknown argument means the source of it cannot be determined. 

There are several types of known arguments, which are: 1) flag arguments, 2) range arguments and 3) distinct value arguments. The flag argument is a combination of multiple flags using the logic OR operation. Each flag has its own meaning. For instance, the flag of the fopen API is combined with different flags (e.g., ``r", ``w", etc.). The range argument is a value with an upper and lower limit. For example, a memory address has its star address and the end address. The distinct value argument may take a specific value of a value set. 

Fortunately, many API arguments are constants or fetched from a determined data set, which can be easily reached through backward data flow analysis. These arguments are usually the access modes, data sizes, protection flags, etc. The obtained API arguments are used to further configure Seccomp with the mapping of API arguments and syscall arguments.

\subsubsection{Backward Taint Analysis} 

To determine the possible value of a syscall argument, a backward taint analysis approach is proposed to generate the data dependency graph of the argument. If all of the dependency variables are constants or API arguments, the data set of the syscall argument is determined. The taint analysis is employed because it can obtain the data dependency relationship without analyzing the complicated constraints in the control flow. In contrast, it is difficult to obtain the value set of a syscall argument from the API entries by symbolic execution. There are many variables that may affect the execution paths, and some of them cannot be analyzed statically. But, the taint analysis can skip these constraints. The taint analysis starts from the syscall arguments instead of API arguments, because we aim to construct the data dependency graph of every syscall argument. Some syscall arguments are assigned by some constants near the syscall invocation functions, which are difficult to analyze by tainting the API arguments. 

The backward taint analysis takes the control flow graph of a library API as the input and outputs the data dependency graphs of every dependent syscall argument. Since Seccomp cannot filter pointers, so the analysis only analyzes the arguments with non-pointer types. For a syscall in the control flow graph of a library API, the non-pointer arguments are set as the taint source. For every argument, the variables that are passed to the tainted variables are set as tainted. If the tainted variable is calculated from other variables, these variables are set as tainted, and the corresponding calculation statements are recorded. When the taint analysis reaches the function arguments, the parent functions in the control flow graph are continued to be analyzed, and the corresponding arguments in the parent functions are set as tainted. The taint sinks when the tainted variable is a constant, an API argument or an object field. By combining the backward data flow propagation process and the recorded statements, the data dependency graph is constructed.

If all of the data resources of a syscall argument are determined, the syscall argument is determined.

The backward taint analysis is inter-procedural, so it suffers from the path explosion problem due to the indirect call analysis. The indirect call analysis is employed during the control flow graph construction, which is an opening problem in static analysis. As a result, the control flow graph is over-approximated. Since our taint analysis follows the control flow graph, the analysis suffers from the path explosion problem. To mitigate this problem, two approaches are applied. First, we leverage  state-of-art indirect call analysis approaches\cite{lu2019does,lu2019detecting} to reduce the false positives. Second, we have optimized the taint analysis process to make it faster by recording the results of the taint analysis. When the taint analysis analyzes a function for the second time, the tainted variables can be directly obtained according to the records. These approaches cannot solve the problem completely, and we leave this problem in future work. But, the impact of this issue is limited in this paper. The over-approximated taint analysis result will make API arguments correspond to more syscall arguments, so that some unused arguments will be allowed. In the worst cases, our approach cannot obtain the mapping between API arguments and syscall arguments. So, this problem can only affect the capability of argument filtering, and the over-approximated results will not affect the normal execution of the target program. But, our system can still resist the exploitation of many CVEs as described in the evaluation.

\subsubsection{Symbolic Execution} 
To further determine the relationship between API arguments and syscall arguments and shrink the possible syscall arguments at best effort, the symbolic execution is employed, which analyzes the functions that contain the possible data sources. The main idea is to obtain the relationship between the function arguments and the in-function constants that could be the data resources of a syscall argument. If the possible constants of a syscall argument are only determined by an API (or function) argument, the relationship can be obtained.

The symbolic execution depends on the intermediate files generated by the gcc compiler. Before starting to find the argument relationship, we first create a control flow graph of each related function. The control flow graph of a function takes a statement as a node, and a jump statement is a directed edge. A directed edge starting from a non-jump statement (such as an assignment statement) points to the instruction following it. A jump statements may emit one or more directed edges (such as goto and switch statements) to point to other statements. Starting from the first statement of the function, we can simulate the execution of the function.

Before the execution starts, we first assign a special symbol to each argument of the target function. The execution process starts from the first statement of the function. When processing an assignment statement, we record the value assigned to the variable. If the variable is used by the conditional statements (e.g., if or switch), the constraints are analyzed to get each branch's condition. If the data assignment of a syscall argument is only determined by the function's arguments, the mapping is constructed. After the execution, if all of the data sources of a syscall are determined by the function arguments, the data set of the syscall argument can be reduced. If a function argument is identified as the decisive factor of a syscall argument, the argument is set as tainted and the backward taint analysis is employed to further analyze the data sources of it. The most complicated case that our system has handled is converting a string to an integer flag. In this case, the flag is only determined by each character of the string, so the mapping between strings and flags can be constructed.

During the symbolic execution, we find that the same function may be analyzed for many times. Some complicated functions will make the analysis extremely slow. To improve the performance, we record the analysis result of every function, so that there is no need to analyze the function again. In addition, when we analyze a branch, we analyze the two branches at the same time to improve the performance. The analysis records the symbolic correspondence between the parameters of each function and its sub-function parameters in memory objects. The symbolic relationship are stored as strings. The static analysis is performed offline with a powerful server and generates configuration files for the embedded devices, so the analysis overhead is acceptable and will not affect the scalability.

The ability to simulate the execution of functions benefits from the simplicity of statements of the intermediate output by gcc. It is very complicated to simulate the execution of C language, but the statements in the intermediate file are simpler. For example, there are several situations in an assignment statement: conversion, calculation and function calls before the assignment. Fortunately, these problems can be addressed by analyzing the intermediate outputs of the compiling, which can present the operations step by step.

The symbolic execution faces many problems, such as path explosion and difficulty to solve some complicated constraints, which make it difficult to apply it to complex programs. In our approach, symbolic execution is applied in dynamically-linked library functions that contain the possible data sources, and in most cases there is only numeric transfer. Although symbolic execution cannot handle particularly complex cases, our application scenario (i.e., argument transfer in library functions) are relatively simple. In this case, the flag is only determined by each character of the string, so the mapping between strings and flags can be constructed. In the worst cases, the mapping between API arguments and syscall arguments cannot be constructed, and this problem can only affect the capability of argument filtering. So, the impact of the problems of symbolic execution is acceptable.

\subsubsection{Seccomp Configuration Generation}

The mapping between API arguments and syscall arguments is added to the API-Syscall mapping constructed in Section \ref{S:static-analysis}. So, the final API-Syscall mapping can be used to find not only an application's dependent syscall set but also the corresponding arguments in a best-effort way. Based on the dependent syscall list and some arguments, Seccomp configuration can be generated. Since the kernel attack surface is reduced by the embedded kernel security module, the introduced overhead will be lower than self-written tools \cite{mishra2018shredder}.

\section{Implementation}\label{s:implementation}
In this section, we will discuss some implementation details. The prototype system proposed in this paper is implemented based on glibc v2.31 and implemented under the ARM structure. 

\subsection{Compilation}
The main idea of the system is to analyze the relationship between functions, including the function call relationship and the argument transfer relationship between functions. However, the source code is not easy to analyze directly due to many factors, such as macros, etc. In order to facilitate the analysis, the gcc compiler is employed to generate semantic dumps during the compilation. When using gcc to compile the glibc source code, in addition to using the flags to add the debugging information and the optimization flags, two flags for generating special files are also used, which are ``-fdump-ipa-cgraph" and ``-fdump-tree-cfg". After the compilation, we can obtain the ``.cfg" file and the ``.cgraph" file corresponding to each ``.c" file. The cfg file uses a simple syntax to describe the functions defined in the c source file, and the cgraph file contains the alias information of the function. With these two files, the function relationship analysis becomes easier.

\subsection{Indirect-call analysis}
Inspired by \cite{lu2019does,lu2019detecting}, the indirect calls are analyzed by the two-level analysis approach, which first identifies the indirect callsites. To this end, the compiling dumps of the dependent libraries are used. These dumps are generated during the compiling, so they contain the semantic information, such as functions, code blocks and statements. If the operand of a callsite is a variable, the callsite is identified as an indirect call.

To identify the address-taken functions, we check if the operand of each statement is a function. Besides, if a function name is used as the argument of another function, the function is also identified. In this step, the mapping between function pointers and objects is constructed. If a function address is stored in an object, a two-tuple (function, object type) will be recorded. This is based on the observation that the targets of indirect calls are usually fetched from objects, and the addresses are stored in the objects previously. So, only pointers stored in objects of the same type can be the targets of indirect calls.

For each indirect call, the possible targets are first identified through the type-based alias analysis from the address-taken functions. The type-based alias analysis compares the number and types of the callsite's arguments with those of address-taken functions. To this end, we need to collect the argument types of every callsite and address-taken function. The function arguments are defined in the function definition. To identify the types of callsite arguments, the definitions of them are analyzed through the backward data analysis. Based on the observation that the arguments are declared in the same function of the callsite, the types of them can be easily collected. By comparing the argument types of callsite and callees, the possible targets can be identified. As discussed above, only pointers stored in objects of the same type can be the targets of indirect calls. So, if the function address is fetched from an object, the type of it is collected. The possible targets are limited to the functions with the same object type in the mapping.

\subsection{Symbolic Execution}
The symbolic execution is performed within functions, and it follows the function control flow. When a function is invoked, the argument information of the function is analyzed. If the argument is a variable name, we replace the variable name with the content of a variable (i.e., a symbol). If the argument is a number, we use the number as the symbol directly. For other types of values, we do not process them for now. For a function, if the value of a certain argument belongs to a finite set of numbers, the value range of the argument is determinable. After we analyze the functions following the control flow graph, we can further analyze the data relationship layer by layer through function calls.

The symbolic execution also runs at best effort, since blocking a syscall argument incorrectly may make the program crash. So, if we find that the value of a syscall argument cannot be determined, the argument will not be restricted.

\section{Evaluation}\label{s:evaluation}
In order to evaluate our prototype system, we select 100 programs from the ARM Linux, which are used to test the effectiveness and performance of our system. To obtain the analysis targets, we have collected 1,208 ARM firmware from the dataset of Firmadyne \cite{chen2016towards}, and leveraged binwalk\cite{binwalk} and Firmwalker\cite{firmwalker} to unzip the file images. From the unzipped file images, we selected 100 common glibc-based programs for the evaluation, which are mainly in the /bin/, /user/bin and /sbin/ file directories. There are 2 criteria for our selection of target programs. First, the target program should be included in most (more than 90\%) images. Second, the program should be written in C language and based on the glibc, since our approach leverages glibc and C programs as representative analysis targets. The average size of the selected programs is 71.7 kB.

\subsection{Effectiveness}
First of all, we analyze whether Seccomp restrictions imposed by our system could affect the normal execution of the target programs. Then, we test the effectiveness of blocking unused syscalls and syscall arguments.

The Seccomp profile generated by our system can block some syscalls and syscall arguments of a target program. However, if one of the dependent syscall or argument is restricted incorrectly, the target program may crash. Therefore, the set of non-disabled syscalls and arguments generated by our system must be over-approximated.

\begin{figure}
    \centering
    \begin{tikzpicture}
    \begin{axis}[
			xlabel = Index of Programs,
			xmin = 0,
			xmax = 100,
			ymin = 0,
			ymax = 250,
			ylabel= Number of Blocked Syscalls and Involved CVEs,
			ymajorgrids = true,
			legend style={
at={(0.5,1.35)},
anchor=north},
			xtick= {0, 10, 20, 30, 40, 50, 60, 70, 80, 90, 100},
			]
    \addplot[blue] plot coordinates {
(1, 36) (2, 36) (3, 36) (4, 36) (5, 36) (6, 36) (7, 36) (8, 36) (9, 36) (10, 36) (11, 36) (12, 36) (13, 34) (14, 36) (15, 36) (16, 36) (17, 35) (18, 36) (19, 35) (20, 35) (21, 36) (22, 36) (23, 36) (24, 36) (25, 36) (26, 36) (27, 36) (28, 36) (29, 36) (30, 36) (31, 36) (32, 36) (33, 34) (34, 36) (35, 36) (36, 36) (37, 36) (38, 36) (39, 36) (40, 36) (41, 36) (42, 36) (43, 36) (44, 36) (45, 36) (46, 36) (47, 36) (48, 35) (49, 36) (50, 36) (51, 36) (52, 36) (53, 35) (54, 36) (55, 36) (56, 36) (57, 36) (58, 36) (59, 35) (60, 36) (61, 31) (62, 36) (63, 36) (64, 36) (65, 36) (66, 36) (67, 36) (68, 36) (69, 36) (70, 36) (71, 35) (72, 36) (73, 36) (74, 36) (75, 36) (76, 36) (77, 36) (78, 36) (79, 36) (80, 36) (81, 35) (82, 36) (83, 34) (84, 36) (85, 36) (86, 36) (87, 36) (88, 36) (89, 36) (90, 36) (91, 36) (92, 36) (93, 36) (94, 36) (95, 35) (96, 36) (97, 36) (98, 35) (99, 36) (100, 36)
    };
    \addlegendentry{\#CVE mitigated by blocking syscalls}
    \addplot[red] plot coordinates {
(1, 97) (2, 97) (3, 97) (4, 97) (5, 97) (6, 97) (7, 97) (8, 97) (9, 97) (10, 97) (11, 97) (12, 97) (13, 96) (14, 97) (15, 97) (16, 97) (17, 97) (18, 97) (19, 97) (20, 97) (21, 97) (22, 97) (23, 97) (24, 97) (25, 97) (26, 97) (27, 97) (28, 97) (29, 97) (30, 97) (31, 97) (32, 97) (33, 96) (34, 97) (35, 97) (36, 97) (37, 97) (38, 97) (39, 97) (40, 97) (41, 97) (42, 97) (43, 97) (44, 97) (45, 97) (46, 97) (47, 97) (48, 97) (49, 97) (50, 97) (51, 97) (52, 97) (53, 97) (54, 97) (55, 97) (56, 97) (57, 97) (58, 97) (59, 97) (60, 97) (61, 97) (62, 97) (63, 97) (64, 97) (65, 97) (66, 97) (67, 97) (68, 97) (69, 97) (70, 97) (71, 97) (72, 97) (73, 97) (74, 97) (75, 97) (76, 97) (77, 97) (78, 97) (79, 97) (80, 97) (81, 97) (82, 97) (83, 97) (84, 97) (85, 97) (86, 97) (87, 97) (88, 97) (89, 97) (90, 97) (91, 97) (92, 97) (93, 97) (94, 97) (95, 97) (96, 97) (97, 97) (98, 97) (99, 97) (100, 97)
    };
    \addlegendentry{\#CVE mitigated by blocking syscall args}
    \addplot[yellow] plot coordinates {
(1, 201) (2, 201) (3, 202) (4, 201) (5, 200) (6, 200) (7, 200) (8, 201) (9, 201) (10, 201) (11, 202) (12, 202) (13, 198) (14, 202) (15, 201) (16, 200) (17, 201) (18, 201) (19, 199) (20, 199) (21, 202) (22, 200) (23, 200) (24, 202) (25, 202) (26, 202) (27, 202) (28, 202) (29, 202) (30, 202) (31, 202) (32, 201) (33, 198) (34, 202) (35, 202) (36, 202) (37, 202) (38, 202) (39, 202) (40, 202) (41, 200) (42, 196) (43, 201) (44, 202) (45, 202) (46, 202) (47, 202) (48, 200) (49, 202) (50, 200) (51, 201) (52, 201) (53, 196) (54, 200) (55, 201) (56, 201) (57, 202) (58, 201) (59, 198) (60, 202) (61, 201) (62, 201) (63, 202) (64, 202) (65, 202) (66, 202) (67, 202) (68, 202) (69, 202) (70, 201) (71, 198) (72, 202) (73, 201) (74, 201) (75, 201) (76, 201) (77, 201) (78, 202) (79, 200) (80, 202) (81, 200) (82, 201) (83, 200) (84, 201) (85, 199) (86, 201) (87, 201) (88, 202) (89, 202) (90, 202) (91, 201) (92, 201) (93, 202) (94, 201) (95, 200) (96, 202) (97, 202) (98, 200) (99, 200) (100, 202)
    };
    \addlegendentry{\#Blocked syscalls without args}
    \addplot[black] plot coordinates {
(1, 242) (2, 242) (3, 242) (4, 242) (5, 242) (6, 242) (7, 242) (8, 242) (9, 242) (10, 242) (11, 241) (12, 242) (13, 240) (14, 242) (15, 242) (16, 242) (17, 242) (18, 242) (19, 242) (20, 241) (21, 242) (22, 242) (23, 242) (24, 241) (25, 242) (26, 242) (27, 242) (28, 242) (29, 242) (30, 242) (31, 242) (32, 242) (33, 240) (34, 242) (35, 242) (36, 242) (37, 242) (38, 242) (39, 242) (40, 242) (41, 242) (42, 240) (43, 242) (44, 242) (45, 242) (46, 242) (47, 242) (48, 242) (49, 242) (50, 241) (51, 242) (52, 242) (53, 242) (54, 242) (55, 242) (56, 242) (57, 242) (58, 242) (59, 241) (60, 242) (61, 242) (62, 242) (63, 242) (64, 242) (65, 242) (66, 241) (67, 242) (68, 242) (69, 242) (70, 240) (71, 242) (72, 242) (73, 242) (74, 242) (75, 242) (76, 242) (77, 242) (78, 242) (79, 240) (80, 242) (81, 242) (82, 242) (83, 241) (84, 242) (85, 242) (86, 242) (87, 242) (88, 242) (89, 242) (90, 242) (91, 242) (92, 241) (93, 242) (94, 242) (95, 242) (96, 242) (97, 242) (98, 242) (99, 242) (100, 242)
    };
    \addlegendentry{\#Blocked syscalls with args}
    \end{axis}
    \end{tikzpicture}
    \caption{The statistics of the syscall dependency analysis on every program, and the number of CVEs involved.}\label{fig:static-result-everyp-program}
\end{figure}
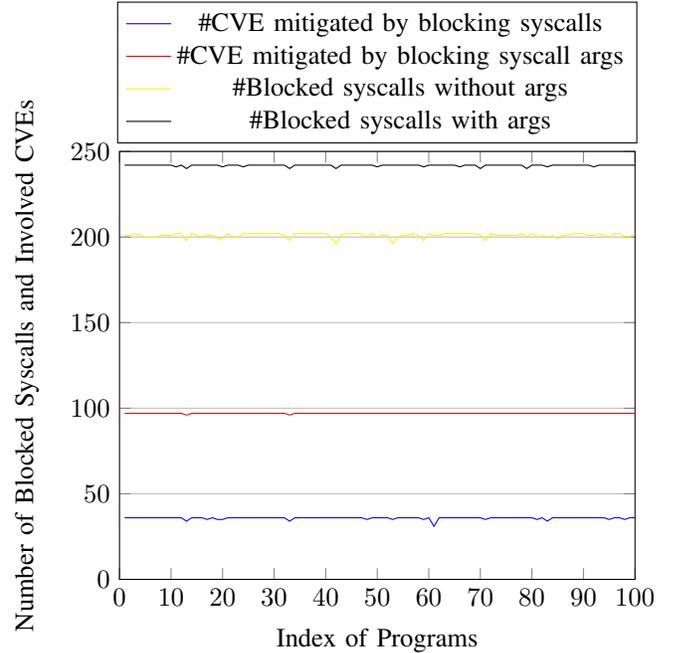

We analyze the target programs to generate Seccomp profiles, including syscall limitation and syscall argument limitation. The experimental results are shown in Figure \ref{fig:static-result-everyp-program}. In the figure, the x-axis is the index of the program, and the y-axis is the number of syscalls or CVEs. In this part, we focus on the yellow and black lines. The yellow line indicates the number of syscalls disabled for the programs, and the black line indicates the number of affected syscalls with arguments. The part between the black line and the yellow line is the number of syscalls whose arguments are restricted. The syscalls involved in the experiment are extracted from the arch-syscall.h header file, which lists a total of 291 syscalls. According to the experimental result, each program has 201 syscalls disabled and 40.82 syscalls with restricted arguments on average, which can significantly reduce the attack surface of kernel.

To verify the correctness of syscall limitation, we leverage the strace to track the execution of each program and record the invoked syscalls of them. All of the invoked syscalls and syscall arguments are within the analysis results, which proves that our system can correctly obtain the over-approximate set of the dependent syscalls and arguments.

\subsection{CVE Mitigation}

We evaluate the system's capability of CVE mitigation. If a syscall that contains a CVE is blocked by our system, the corresponding CVE is also mitigated. So, we collect the mapping between CVEs and syscalls from cve.org to test the capability of CVE mitigation. We check the descriptions of CVEs from 2020 to 2013, and filter out the CVEs whose descriptions clearly contain the keywords "syscall" or "system call". After that, we further select those CVEs that clearly indicate the affected syscalls and arguments, which number is 120. Based on the results, the mapping can be constructed. 

Next, the CVEs that can be mitigated for each program can be analyzed. For each collected CVE, the corresponding syscalls and arguments can be obtained. Then, we check the allowlist of each program to find out if the CVE-related syscalls or arguments are blocked by the Seccomp configuration. If the CVE-related syscalls or arguments can be blocked, the syscalls, arguments and the corresponding CVE are recorded. The results are shown in Figure \ref{fig:static-result-everyp-program}.

As shown in Figure \ref{fig:static-result-everyp-program}, the red line is the number of CVE mitigation by blocking syscalls and syscall arguments, and the blue line is the number of CVE mitigation by only blocking syscalls. The part between the blue line and the red line is the number of newly mitigated CVEs by limiting the syscall arguments. On average, 35.79 CVEs are mitigated by blocking syscalls for one program, and 61.19 CVEs are newly mitigated by restricting syscall arguments. 

From the results, we can find that blocking the accessible syscalls can effectively mitigate CVEs, and further limiting the syscall arguments can mitigate more CVEs. In addition, the number of new CVEs affected by limiting the arguments is larger than that of only blocking syscalls, which shows that further limiting syscall arguments is very meaningful. 

Table \ref{T:cve} illustrates the top 20 syscalls with the related CVEs, which are blocked or restricted with arguments by our system for the target programs. From the table, we can find that the network-related syscalls introduce the most CVEs. There are 120 CVEs in the CVE collection. After we combine the CVE collections mitigated by different programs, 97 of them can be mitigated.

\begin{table*}
  \centering
  \caption{Top 20 syscalls and the related CVEs mitigated by sysverify.}\label{T:cve}
  \begin{tabular}{ccccc}
     \hline
     Syscalls & Number of CVEs & Representative CVEs \\
	 \hline
	recvmsg & 29 & CVE-2013-3228,CVE-2013-3225,CVE-2013-7267 \\
	recvfrom & 29 & CVE-2013-3224,CVE-2013-3223,CVE-2016-10229 \\
	socket & 18 & CVE-2016-10200,CVE-2017-7277,CVE-2013-7339 \\
	setsockopt & 11 & CVE-2017-6074,CVE-2017-6346,CVE-2017-16939 \\
	bind & 9 & CVE-2016-10200,CVE-2017-7277,CVE-2013-7339 \\
	sendmsg & 5 & CVE-2017-9242,CVE-2016-3841,CVE-2018-1130 \\
	mmap & 5 & CVE-2017-14497,CVE-2018-7740,CVE-2016-4794 \\
	ioctl & 5 & CVE-2020-10942,CVE-2013-2239,CVE-2017-18257 \\
	sendto & 5 & CVE-2017-15115,CVE-2017-7308,CVE-2015-2686 \\
	ptrace & 5 & CVE-2013-0871,CVE-2014-3534,CVE-2014-4699 \\
	write & 3 & CVE-2017-7277,CVE-2017-7495,CVE-2015-8019 \\
	accept & 3 & CVE-2015-8970,CVE-2017-8890,CVE-2017-9075 \\
	writev & 3 & CVE-2014-0069,CVE-2016-9755,CVE-2015-8785 \\
	getsockopt & 3 & CVE-2017-15115,CVE-2013-1828,CVE-2013-4588 \\
	connect & 3 & CVE-2017-7277,CVE-2016-9755,CVE-2017-8824 \\
	read & 2 & CVE-2017-7495,CVE-2013-6432 \\
	unshare & 2 & CVE-2017-15115,CVE-2014-7975 \\
	madvise & 2 & CVE-2015-7312,CVE-2014-8173 \\
	syncfs & 2 & CVE-2019-19448,CVE-2019-19813 \\
	brk & 1 & CVE-2020-9391 \\
	 \hline
   \end{tabular}
\end{table*}

Taking the CVE-2017-7308 as a case study, where PoC is given in exploit-db\cite{exploit-db}. The vulnerability locates in the packet\_set\_ring function in the net/packet/af\_packet.c file. When the packet\_set\_ring function encounters the TPACKET\_V3 flag, it will check the size of the private area (i.e., tp\_sizeof\_priv member variable) of the ring buffer block set in the user request. However, the subtraction of unsigned numbers is used in the check and the result is forcibly converted to the int type, which leads to some huge tp\_sizeof\_priv values that can bypass this check. By carefully setting the value of tp\_sizeof\_priv, the attackers can cause out-of-bounds writes to the kernel heap when receiving data packets. Besides, the offset of writing can be controlled, and the KASLR can be bypassed. After that, the SMAP and SMEP are disabled and the kernel heap is reshaped, so that the block of the ring buffer could overwrite the packet\_sock structure and the function pointer field in it can be overwritten. Finally, the commit\_creds function is executed in the context of the user process, and the root privileges can be obtained. The attack vector dependents on two syscalls (i.e., socket and setsocketopt) and the AF\_PACKET flag. If a program does not rely on this flag in normal execution, the flag can be blocked by Seccomp, so that the vulnerability can be mitigated. Shredder\cite{mishra2018shredder} can analyze the library API arguments that may be used by an application. By using Shredder, the dependent APIs and corresponding arguments of a web-related application can be obtained. After that, the corresponding syscall arguments can be determined and blocked by our system. If an application (such as telnet) does not rely on the socket syscall with the AF\_PACKET argument, the argument can be limited by Seccomp. Even if the application is hijacked by an attacker in some way, the attacker cannot exploit this CVE to crash the system or gain root privileges.

Through our observation, some syscalls are easy to restrict, but others are difficult. For instance, the network-related syscalls, such as sockets, have a very short call chain from the related API to this syscall. Many arguments of the API are directly passed to syscalls for use. For this kind of syscalls, even if they cannot be blocked, there is a high possibility to limit their arguments. And, the CVEs related to these syscalls are also easy to mitigate. However, there are many call chains from APIs to syscalls that are very long, which makes it difficult to determine the arguments of syscalls. In this case, if the program uses the API, the corresponding syscall arguments cannot be blocked, so the CVE related to the syscall cannot be mitigated. Some syscall arguments dependents on global variables/pointers in the code implementation, which cannot be handled currently. In addition, if the syscall arguments dependent on complicated data structures, the data source cannot be analyzed by our system for now. And, we leave these problems in the future work.

\subsection{Performance}\label{S:performance}
Our system leverages Seccomp to secure system kernel, the main factor that affects the performance of the target programs is Seccomp. Seccomp is an embedded module in Linux operating system. Seccomp only works when a program configures Seccomp and invokes syscalls. If a program does not configure Seccomp policies, the performance of syscalls will not be affected. So, Seccomp can only introduces overhead to programs that are configured with Seccomp. In this section, we test the performance impact of it. The target programs used in the test come from the lmbench performance test tools. We test the performance from three aspects: memory read and write, file read and TCP bandwidth. 

We use the API provided by the libseccomp library to configure Seccomp for the target programs and adopt a whitelist strategy. The Seccomp rules are automatically generated by our system. For example, if our system finds that the target program may use the socket syscall, and the first argument can only be AF\_INET, then the system will generate a corresponding Seccomp rule. This rule will enable the program to successfully pass the check when calling the socket syscall with the AF\_INET flag as the first argument. The LMBench is selected as the benchmark to test the performance. For comparison, we also test the performance of the approaches that only limit syscalls (e.g., Confine\cite{ghavamnia2020confine}) with the same testing tools and methods.

\begin{table*}
  \caption{Performance Comparison in three cases: no limitation, limiting syscalls, limiting syscalls \& parameters.}
  \centering
  \label{T:cost}
  \begin{tabular}{cccccc}
	\hline
	\textbf{Benchmark}&\textbf{Block/packet size}&\textbf{Benchmark Mode}&\textbf{No limitation}&\textbf{Limiting	syscalls}&\textbf{Limiting syscalls \& params}\\
	\hline
	\multirow{21}{*}{bw\_mem} & \multirow{6}{*}{32k} & bcopy & 44265.02 & 44257.37 & 44240.37\\
	& & bzero & 88737.19 & 88674.75 & 88693.81\\
	& & fcp & 11167.81 & 11175.11 & 11173.21\\
	& & frd & 13375.55 & 13365.82 & 13371.43\\
	& & fwr & 22309.37 & 22306.72 & 22289.05\\
	& & rdwr & 44567.34 & 44559.24 & 44567.34\\
	\cline{2-6}
	& \multirow{6}{*}{64k} & bcopy & 43674.58 & 43694.70 & 43658.22\\
	& & bzero & 89058.38 & 88953.72 & 89080.88\\
	& & fcp & 11125.16 & 11124.86 & 11122.48\\
	& & frd & 13361.36 & 13366.21 & 13375.91\\
	& & fwr & 22310.13 & 22322.05 & 22328.74\\
	& & rdwr & 43960.26 & 43971.15 & 43960.26\\
	\cline{2-6}
	& \multirow{6}{*}{128k} & bcopy & 39128.11 & 39302.49 & 39326.51\\
	& & bzero & 86338.05 & 88660.29 & 88471.23\\
	& & fcp & 11033.64 & 11027.64 & 11024.30\\
	& & frd & 13214.15 & 13213.55 & 13221.13\\
	& & fwr & 22336.87 & 22309.62 & 22316.70\\
	& & rdwr & 30929.77 & 31001.92 & 30981.99\\
	\cline{1-6}
	\multirow{15}{*}{bw\_file\_rd} & \multirow{2}{*}{128k} & io\_only & 9235.10 & 9104.86 & 9149.07\\
	& & open2close & 8465.41 & 8375.36 & 8384.34\\
	\cline{2-6}
	& \multirow{2}{*}{256k} & io\_only & 9216.76 & 9173.34 & 9165.17\\
	& & open2close & 8851.62 & 8742.99 & 8741.35\\
	\cline{2-6}
	& \multirow{2}{*}{512k} & io\_only & 8639.44 & 8567.75 & 8687.13\\
	& & open2close & 8352.87 & 8325.10 & 8304.56\\
	\cline{2-6}
	& \multirow{2}{*}{1m} & io\_only & 8140.52 & 8106.05 & 8122.64\\
	& & open2close & 7983.59 & 7929.11 & 7964.36\\
	\cline{2-6}
	& \multirow{2}{*}{2m} & io\_only & 7867.13 & 7850.30 & 7751.44\\
	& & open2close & 7666.43 & 7649.65 & 7660.83\\
	\cline{1-6}
	\multirow{5}{*}{bw\_tcp} & 128 & $\backslash$ & 343.79 & 350.01 & 357.38\\
	& 256 & $\backslash$ & 666.54 & 662.68 & 678.05\\
	& 512 & $\backslash$ & 1139.12 & 1230.59 & 1250.85\\
	& 1024 & $\backslash$ & 1953.47 & 2075.29 & 2106.87\\
	& 1437 & $\backslash$ & 2343.87 & 2572.09 & 2610.74\\
	\cline{1-6}
  \end{tabular}
  \footnotesize
 \\
\end{table*}

The results of the benchmark with different security policies (e.g., without Seccomp, with Seccomp limiting syscalls, with Seccomp limiting syscalls and arguments) are shown in Table \ref{T:cost}. To analyze the performance impacts, we calculate the average performance degradation on various metrics by limiting only syscalls and limiting syscalls and arguments. According to the calculation results, in the case of only limiting syscalls, the average performance of bw\_mem decreases by 0.23\%, bw\_file\_rd decreases by 0.69\% on average, and bw\_tcp decreases by 5.28\% on average. When limiting both syscalls and arguments, there are similar results, 0.22\%, 0.68\%, and 6.95\%, respectively. From the results, we can find that our approach introduces less than 2\% more performance degradation than other Seccomp-based approaches that only limit syscalls. In some cases, the performance with Seccomp limitation is higher than that without Seccomp limitation, which means the performance fluctuation is higher than the performance impact of Seccomp limitation.

The performance can be further optimized by changing the order of Seccomp rules. After enabling Seccomp, an invoked syscall is matched with each Seccomp rule in turn until the match is successful or all rules are not matched. So, the order of the Seccomp rules is important for the performance. If the order of Seccomp rules is not appropriate, it is possible that every syscall invoked by the target program will be matched for a long time. So, the order of the rules can be adjusted to further reduce the performance loss. A straightforward way is to use the strace to track the execution of the program, and rank the syscalls that are used frequently. Based on this order, the performance of Seccomp can be optimized.

\section{Related Work}\label{s:related-work}
\subsection{Lightweight VM-based Containers}
It is a major way to leverage the lightweight virtual machines (VMs) to isolate applications, because VMs are considered to be more secure than containers and sandboxes \cite{combe2016docker}. Many lightweight VMs are proposed to protect the operating systems from applications. The unikernel \cite{madhavapeddy2014unikernels} compiles the applications with the library OS\cite{tsai2014cooperation} into a lightweight VM image, so that the image can be executed as a lightweight VM. Since the OS is integrated into the applications, the OS is not isolated from the user space, which is not secure. In addition, all of the applications should be recompiled or redesigned. The Kata Container \cite{kata} leverages VMs to isolate containers by running a container in a lightweight VM, which starts fast and consumes less resources. The goal of the Kata Container is to improve the isolation of the containers to the level of the virtual machines while maintaining the performance of the containers. However, the performance of it is lower than that of a Docker container. The gVisor\cite{gvisor} leverages the para-virtualization to isolate containers, which has two parts: Sentry and Gofer. Sentry emulates a virtual kernel for containers, which can handle most of syscalls invoked by containers. Therefore, the attack surface of the kernel is reduced. Gofer can redirect the I/O requests of containers to the host operating system. Compared with the Kata Container, the isolation of Gvisor is weaker, since some syscalls are handled by the host operating system. 

In summary, a big problem of applying VM-based isolation approaches in IoT scenario is that most of the IoT/embedded devices do not have enough resources to support VMs, so some lightweight security enhancement approaches are needed.

\subsection{Securing Operating Systems}
The operating system is not secure in multi-application systems. Bastion \cite{nam2020bastion} proves the weaknesses of the Linux network isolation, and performs cross-process attacks on the Linux system. \cite{gao2017containerleaks,gao2018study} show the weakness of the isolation of the proc file system. By using it, a process can obtain the information of other processes on the same host.

To secure the operating systems for multi-application systems, reducing the attack surface of the operating system is an important way. \cite{lei2017speaker,wan2017mining,barlev2016secure,ghavamnia2020confine} filter the unused syscalls for user-space applications, so that it is difficult for user-space applications to exploit the kernel vulnerabilities. Based on the observation that an application usually invokes different syscalls in different execution stages (e.g., initialization stage, servicing stage and exiting stage), \cite{ghavamnia2020temporal} blocks different syscalls according to the application’s execution stages. To this end, it identifies the dependent syscalls of different running stages. But, these approaches are coarse-grained, which can only limit the syscall numbers. In addition, these systems are designed for cloud services, and some of them introduce additional security monitors in the target systems. Unlike these systems, SCONE\cite{arnautov2016scone} compiles the dependent system service code into the applications by leveraging the libOS, so that it does not invoke syscalls. However, the recompilation makes the approach's versatility worse.

\subsection{Program Debloating}
Besides blocking the syscalls, removing the unused code from the applications and the dependent libraries is another way to secure the operating systems. Piece-wise debloating \cite{quach2018debloating} only loads the necessary dependent libraries and replaces the unnecessary parts with null code. Nibbler\cite{agadakos2019nibbler} removes the unused code by analyzing the function call graph of the target application. LibFilter\cite{shteinfeldlibfilter} removes the unused functions in the dependent libraries. In contrast, Razor\cite{qian2019razor} constructs the function call graph through dynamic tracking. However, these the program debloating approaches need to redesign the applications or dynamically-linked libraries, so it is not easy to apply the approaches to commercial systems widely.

\subsection{IoT Security}
IoT security 
\cite{khan2018iot,hassija2019survey,hassan2019current,xiao2018iot,al2020survey} is a hot topic, which is related with device security\cite{7958583,7163056}, network security\cite{huang2019not}, cloud security\cite{zhan2018high,zhou2019discovering}, etc. This paper focuses on the system security of IoT devices, so we mainly discuss the IoT device security.

IoT device firmware is usually developed by low-level languages (such as assembly and C language), so coding or design bugs are inevitably introduced in the development process. Due to limited hardware resources, IoT devices usually lack necessary dynamic system defense measures such as CFI (Control Flow Integrity), etc., which make attackers exploit vulnerabilities more easily. A number of studies have pointed out that the memory vulnerabilities caused by code injection attacks are common in firmware \cite{7958583,7163056}, and hijacking the control flow of IoT applications through modifying the function return address is still a main threat\cite{almakhdhub2020mu,zhou2020silhouette}. 

The limited software and hardware resources of IoT devices are one of the main reasons why IoT devices are more vulnerable. Therefore, many researches focus on securing IoT systems. 

The main approach is to protect the integrity of the program control flow, which can ensure the integrity of the function return address to deal with control flow hijacking attacks. $\mu$RAI\cite{almakhdhub2020mu} stores the effective return address collection of the functions in the non-writable memory area, and makes the function return addresses fetched from the collection to ensure that the return address will not be tampered with. Silhouette\cite{zhou2020silhouette} is based on the ``Shadow Stack" technology to defend against control flow hijacking attacks. It configures a memory protection unit to implement memory access rules to ensure that the program must return to a legal target address after the return instruction is executed. However, these approaches introduce runtime overhead to the target applications. In addition, the OS is not protected when the applications are compromised.

Many firmware do not have operating systems. To overcome the problems that there are no system protections, some studies divide the firmware into different components to implement least-privileged isolation. EPOXY\cite{7958583} identifies instructions that require high privileges through static analysis, and then provides stack protection, code and data area isolation. But it cannot perform process-level code isolation. MINION\cite{kim2018securing} leverages the MPU to replace the memory area accessible by the process during context switching, thus realizing the isolation of the process memory space. But, the method to divide the data and code areas cannot be applied to complicated applications. ACES\cite{clements2018aces} overcomes the shortcomings of the above two schemes. By automatically identifying and separating the minimum execution unit of the firmware, a more fine-grained authority can be identified. 

\section{Conclusion}\label{s:conclusion}
This paper proposes a novel static dependent syscall analysis approach for IoT applications, which can obtain the dependent syscalls and the corresponding arguments. So that, a fine-grained kernel access limitation can be performed for the IoT applications. To this end, the mapping between dynamic library APIs and syscalls is built through static analysis. A novel argument mapping construction approach based on backward taint analysis and symbolic execution is proposed to further map the arguments between APIs and syscalls. After obtaining the dependent library APIs and the corresponding arguments of an application, the fine-grained kernel access control policy can be generated. The experimental results show that the system can block the unused syscalls and arguments from the target programs and mitigate the CVEs included in the corresponding syscalls with acceptable overhead. In future work, we are going to apply and test our system in other computing systems, such as containers.

\section*{Acknowledgments}

This work was supported by the National Key R\&D Program of China (No.2021YFB2012402) and the National Natural Science Foundation of China under Grants NO. 61872111.

\bibliographystyle{IEEEtran}
\bibliography{sec}

\end{document}